\documentclass[runningheads, a4paper]{llncs}
\usepackage{amssymb}
\usepackage{graphicx}
\usepackage{url}
\usepackage{algorithmic}
\usepackage{algorithm}
\usepackage{booktabs}
\usepackage{arydshln}

\newcommand{\SWITCH}[1]{\STATE \textbf{switch} (#1)}
\newcommand{\ENDSWITCH}{\STATE \textbf{end switch}}
\newcommand{\BREAK}{\STATE \textbf{break}}
\newcommand{\CASE}[1]{\STATE \textbf{case} #1\textbf{:} \begin{ALC@g}}
\newcommand{\ENDCASE}{\end{ALC@g}}

\newcommand{\DEFAULT}{\STATE \textbf{default:} \begin{ALC@g}}
\newcommand{\ENDDEFAULT}{\end{ALC@g}}
\newcommand{\DEFAULTLINE}[1]{\STATE \textbf{default:} }

\title{Recollection: an Alternative Restoration Technique for Constraint Programming Systems}

\begin{document}

\author{Yong Lin and Martin Henz}

\titlerunning{Recollection}

\institute{
School of Computing, National University of Singapore, Singapore\\
\email{\{linyong, henz\}@comp.nus.edu.sg}
}

\date{}
\maketitle

\begin{abstract}

Search is a key service within constraint programming systems, and it demands the restoration of previously accessed states during the exploration of a search tree. Restoration proceeds either bottom-up within the tree to roll back previously performed operations using a trail, or top-down to redo them, starting from a previously stored state and using suitable information stored along the way. In this paper, we elucidate existing restoration techniques using a pair
of abstract methods and employ them to present a new technique that we call recollection. The proposed technique stores the variables that were affected by constraint propagation during fix points reasoning steps, and it conducts neither operation roll-back nor recomputation, while consuming much less memory than storing previous visited states. We implemented this idea as a prototype within the Gecode solver. An empirical evaluation reveals that constraint problems with expensive propagation and frequent failures can benefit from recollection with respect to runtime at the expense of a marginal increase in memory consumption, comparing with the most competitive variant of recomputation.

\end{abstract}

\section{Introduction}

Constraint programming is a tool for modelling and solving constraint satisfaction problems. To support this tool, constraint programming systems provide a set of services to conduct constraint-based search. Among these services, restoration implements the process of restoring a previously reached computation state. A straightforward albeit memory-intensive form of restoration is afforded by making a \emph{copy} of each previously visited state; restoration then just retrieves the expected one from memory. Less memory-intensive alternatives construct the required state based on suitable \emph{chunks} of information collected during search. This reconstruction can be achieved by either \emph{trailing} or \emph{recomputation}. 

% Trailing would \emph{undo} the changes, and
% recomputation will \emph{redo} previously accomplished computing work.

%Constraint programming (CP) systems provide a set of components to support constraint-based search, an effective tool for solving many combinatorial search and optimization problems.  Of all components in CP systems, restoration provides the service of restoring previously reached computation states.

Implementations of logic programming (Prolog), an ancestor of constraint programming, use trailing as the main restoration technique. Most of today's CP systems are constraint logic programming~\cite{clp} systems that evolved from Prolog and inherited its trailing-based restoration paradigm, such as
ECL$^{i}$PS$^{e}$~\cite{eclipse}, CHIP~\cite{CHIP}, clp(FD)~\cite{clpfd} and SICStus Prolog~\cite{journals/tplp/CarlssonM12}. This technique rolls back previously performed operations to restore an earlier state by utilizing undo information maintained in a trail structure, proceeding in a \emph{bottom-up}
direction within the search tree. By contrast, recomputation restores a state by redoing previous computation, which follows a \emph{top-down} manner. Recomputation, combined with copying in various ways, is the restoration technique underlying the concurrent constraint programming system Mozart/Oz~\cite{ozmodel}, as well as the constraint programming solver Gecode~\cite{Gecode}.

This paper investigates a top-down restoration technique that we call \emph{recollection}, which avoids redoing computation, while consuming much less memory than copying. Its central idea is simple; we incrementally memoize the variable domains that were updated to reason fix point state for later restoration.

\paragraph{Plan of the paper.}
The following Section~\ref{cp} introduces some basic notions referred to in the rest of the paper. Section~\ref{restoration} briefly reviews the current available restoration techniques using a pair of abstract methods, and Section~\ref{recollection} presents our proposed \emph{recollection} technique together with its
implementation issues. Section~\ref{evaluation} describes the evaluation of our implemented recollection on selected benchmark problems, and we conclude in Section~\ref{conclusion}.

\section{Constraint-based Problem Solving} \label{cp}
%\input{preliminary}
%

%Formally, a constraint satisfaction problems is defined by a sequence of variables $V_{1}$, $V_{2}$, $\ldots$, $V_{n}$  and a collection of constraints $C_{1}$, $C_{2}$, $\ldots$, $C_{m}$. Each variable $V_{i}$ has a domain $D_{i}$ which maps it to a finite set of integers. A variable $V_{i}$ is \emph{fixed} by a domain if $\|V_{i}\|$ = 1 while a variable is \emph{false} if $\|V_{i}\|$ = 0. A \emph{solution} is identified provided every variable is fixed and an \emph{inconsistency} occurs if any variable is false. A constraint $C_{k}$ posts on a subset of variables and specifies the values of variables that are not conflicting. \emph{Constraint propagation} shrinks variable domains through propagating the implications of constraints to variables.

This section introduces the basic terms and reviews the main principles of constraint programming to provide the context for the rest part of this paper.

\subsection{Basics}

A constraint satisfaction problem defines a finite set of variables together with constraints. Each variable has a domain to specify the finite set of values it can take, and a variable is \emph{fixed} if its domain is a singleton set. The conjunction of variables forms a \emph{store}, which maps variables to their domains. A constraint describes a certain relationship over a subset of variables to restrict their value combinations that can appear in the solutions to the problem. In constraint programming systems, a constraint is implemented as a \emph{propagator}, and propagators amplify their store by executing their filtering algorithms to rule out inconsistent values from variable domains. 

A store and its connected propagators form a \emph{state}, and the store provides a communication channel for its propagators. Specifically, variable domain changes by running a propagator will be reflected in the store, which may trigger the execution of other propagators to filter more values from other variable domains. This process is called \emph{constraint propagation} (propagation for short).  A propagation is \emph{strong} if it is able to change the majority of variables within the problem; otherwise, the propagation is \emph{weak}. During propagation, if any variable's domain becomes empty, the engine infers an \emph{inconsistency} and the current state becomes \emph{failed}; if all variables become fixed, the state is in a \emph{solved} status and represents a \emph{solution}. Nevertheless, constraint propagation alone is generally insufficient to identify a solution; it will reach a \emph{fix point} when constraint propagation cannot make further changes to the store.

In such a circumstance, search is required and it equivalently splits the fix point state into multiple further constrained states. The split is called \emph{branching} and it is achieved by invoking a \emph{brancher} in a constraint programming system. The invocation of a brancher generates a \emph{choice} of the fix point state, and the choice includes multiple  mutually exclusive constraint alternatives\footnote{We restrict our discussion to binary choice in this work.}. The search process \emph{commits} an alternative of the choice to visit the corresponding further constrained state, and constraint propagation can resume at the further constrained state. A choice is \emph{open} if it has an uncommitted alternative; otherwise, it is \emph{closed}.

%To proceed, the fix point problem will be equivalently split into further constrained subproblems, and this task is called \emph{branching} and performed by a \textit{brancher} in a CP system. A brancher takes a fix point as input and produces a \emph{choice} which includes a number of constraint \emph{alternatives}; these alternatives can be respectively committed to the problem to direct exploration. A choice is \emph{open} if it has at least one uncommitted alternative. The search procedure commits alternatives to current states following a certain strategy (e.g. DFS) and then resumes constraint propagation. These steps repeat until ultimately a solution (or the optimal solution etc.) is found or the problem is proved to be non-solvable.

\subsection{Search}

Search is a central service in constraint programming systems and it is programmed as a \emph{search engine}. The search consists of \emph{exploration} and \emph{restoration}. Exploration alternates between constraint propagation and branching, and this leads to a tree of states, the \emph{search tree}. In such a search tree, branches represent constraints, internal nodes are fix points and the leaf nodes are either solved or inconsistent states. Inconsistency indicates a false search direction, and restoration needs to recover a previously accessed internal state, \emph{target state}, to guide exploration to other part of the search tree. An important step for restoration is to decide which state to restore after encountering an inconsistency, and in this paper, we focus on \emph{chronological backtracking}. 
 
In the search tree, the \emph{root} state is the fix point state reasoned on the original problem, and the \emph{current} state is the one that search engine is interacting. The branches and fix points between the \emph{root} and \emph{current} node form a \emph{path}\footnote{We focus on sequential search in this paper and therefore there is a single path.}. In a constraint programming system, the path is designed to store the information that has been utilized to reach current state from root, and the system makes use of this information to conduct restoration.  

%The path should store sufficient information as search proceeds to guarantee that the required node can be restored when exploration encounters an inconsistency. To facilitate the discussion of restoration, we focus our attention on \emph{chronological backtracking} in this paper. 
  
%
 \begin{algorithm}[h!]
\caption{Depth First Search \label{alg1}}
 \algsetup{linenosize=\tiny}
  \scriptsize
\begin{algorithmic}[1]
   \REQUIRE State \emph{S}, Stack \emph{ST}
   \ENSURE  Solution  State
 % \STATE Stack \emph{ST} $\leftarrow$ $\emptyset$
  \WHILE{\textbf{\textit{true}}}
   %\STATE prop\_result $\Leftarrow$
%   \SWITCH {\emph{\textbf{Propagate}}(\emph{S})}
   \STATE Log \emph{log}
   \SWITCH {Propagate(\emph{S}, \emph{log})}

   \CASE {\textbf{\textit{solved}}}
       \RETURN  \emph{S}
     \ENDCASE

    \CASE { \textbf{\textit{inconsistency}}}
       \STATE  \emph{S} $\leftarrow$ Restore(\emph{S}, \emph{ST},  \emph{log})
       \IF { \emph{S} $=$ NULL} \RETURN \textbf{\emph{non-solvable}} \ENDIF
       \STATE Chunk \emph{chunk} $\leftarrow$ getTop(\emph{ST})
       \STATE  Choice \emph{choice} $\leftarrow$ getChoice(\emph{chunk})
      % \STATE  nextAlt(\emph{choice'})
       \STATE  Commit(\emph{S}, \emph{choice}, \textbf{\textbf{\emph{second}}})
       \BREAK
     \ENDCASE

     \CASE {\textbf{\textit{fix\_point}}}
       \STATE Choice \emph{choice} $\leftarrow$ Branch(\emph{S})
        \STATE Commit(\emph{S}, \emph{choice}, \emph{\textbf{first}})
        \STATE Chunk \emph{chunk} $\leftarrow$ Record(\emph{S}, \emph{choice}, \emph{log})
        \STATE  Push(\emph{ST}, \emph{chunk})
        \STATE \emph{log} $\leftarrow$ $\emptyset$
     \ENDCASE

     \ENDSWITCH
    \ENDWHILE
\end{algorithmic}
\end{algorithm}

 Algorithm~\ref{alg1} describes a DFS search without committing to a particular restoration technique. In this pseudo-code, the variable \emph{S} refers current state; the path information is maintained explicitly using a stack~\emph{ST}. The method $\textit{Propagate}()$ (Line 3) conducts constraint propagation within state \emph{S} and the \emph{log} records the changes enforced by the propagation to state \emph{S}. A switch statement responds according to the propagation result. Specifically, if a state is solved, it will be returned as a solution.  If propagation exhibits an inconsistency, the search engine will call the method \emph{Restore} to restore a state, from which to explore the second branch of the node. If a fix point is reasoned, the search engine will call Branch() to generate a choice of the state and then commits it to the first alternative by the Commit() method; subsequently, a chunk will be constructed (Line 18) and pushed onto the stack  \emph{ST}. 
 
In the pseudo-code, the methods \textit{Record} and \textit{Restore} form a pair of abstract methods, whose implementations determine the way to conduct restoration. In the following sections, we discuss restoration techniques by describing this pair of methods.

\section{Restoration} \label{restoration}

The restoration of a previously accessed state can be achieved by either reconstruction or memorization. Memorization has been implemented by copying (Section~\ref{copying}); reconstruction can be approached by trailing (Section~\ref{trailing}) and recomputation (Section~\ref{recomp}). In this section, we briefly go through these restoration techniques together with their characteristics.

\subsection{Trailing} \label{trailing}

 Trailing-based systems implement the $\textit{Record}$ method to accumulate operation undo information in a trail structure (it is the stack \emph{ST} in this context). Conceptually, the undo information is expected to describe the changes enforced to the state. In practical implementations, systems mostly prefer maintaining the original images before updates. Examples are single-value~\cite{abstractMachine}, Time-Stamping and Multiple-Value trail (see~\cite{trailImp}) etc. For a comprehensive description, please refer to~\cite{clpSurvey}.

The updating of a trail structure interleaves with propagation since variables keep changing during propagation. In Algorithm \ref{alg1}, a global data structure \emph{log} is introduced to collect undo information. If propagation leads to a fix point, the collected undo information will be wrapped into the chunk and pushed onto~\emph{ST}. Algorithm~\ref{alg3} shows that trailing-based restoration first cancels the operations stored in the \emph{log}. Subsequently, the chunks in \emph{ST} are accessed one by one to conduct a step-wise roll-back; this process iterates until it backtracks to a state that owns an open choice.
  
\begin{algorithm}[h!]
\caption{$\textit{Restore}_{trail}$ \label{alg3} }
 \algsetup{linenosize=\tiny}
  \scriptsize
\begin{algorithmic}[1]
   \REQUIRE State \emph{S}, Stack \emph{ST}, Log \emph{log}
   \ENSURE State \emph{S}
    \STATE undo(\emph{S}, \emph{log})
    \STATE Chunk \emph{chunk} $\leftarrow$ getTop(\emph{ST})
    \STATE Choice \emph{choice} $\leftarrow$ getChoice(\emph{chunk})
   \WHILE {\emph{choice} has no uncommitted alternative}
     \STATE \emph{log} $\leftarrow$ getLog(\emph{chunk})
     \STATE undo(\emph{S}, \emph{log})
     \STATE Pop(\emph{ST})    
     \IF {Size(\emph{ST}) = 0 } \RETURN NULL \ENDIF 
     %\STATE Assert(! isEmpty(\emph{ST}) )
     \STATE chunk $\leftarrow$ getTop(\emph{ST})
     \STATE \emph{choice} $\leftarrow$ getChoice(\emph{chunk})
   \ENDWHILE
 \RETURN \emph{S}
\end{algorithmic}
\end{algorithm}

\begin{algorithm} [h!]
\caption{EXPOSE\_OPEN\_CHOICE \label{alg0}}
 \algsetup{linenosize=\tiny}
  \scriptsize
\begin{algorithmic}[1]
   \REQUIRE Stack \emph{ST}
   \ENSURE  Stack \emph{ST}

   \STATE Chunk \emph{chunk} $\leftarrow$ getTop(\emph{ST})
   \STATE Choice \emph{choice} $\leftarrow$ getChoice(\emph{chunk})
   \WHILE{\emph{choice} has no uncommitted alternative}
   \STATE  Pop(\emph{ST})
   \IF { Size(\emph{ST}) = 0 } \RETURN NULL \ENDIF
   \STATE  \emph{chunk}  $\leftarrow$ getTop(\emph{ST})
   \STATE \emph{choice} $\leftarrow$ getChoice(\emph{chunk})
   \ENDWHILE

   \RETURN \emph{ST}
\end{algorithmic}
\end{algorithm}

\subsection{Copying} \label{copying}

Copying-based state restoration defines the $\textit{Record}_{copy}$ method to store a copy of the entire state in each constructed chunk. The corresponding $\textit{Restore}_{copy}$ method is straightforward: retrieve the chunk containing the expected state and return. For later reuse, the function EXPOSE\_OPEN\_CHOICE in Algorithm~\ref{alg0} implements chronological backtracking by popping chunks from the stack until it recognizes a chunk that contains an open choice.

\begin{algorithm}[h!]
\caption{ $\textit{Restore}_{copy}$ \label{algcopy} }
 \algsetup{linenosize=\tiny}
  \scriptsize
\begin{algorithmic}[1]
   \REQUIRE State \emph{S}, Stack \emph{ST}, Log \emph{log} (ignored)
   \ENSURE State \emph{S}
   \STATE \textbf{delete} \emph{S}
     \STATE \textbf{call} EXPOSE\_OPEN\_CHOICE
     \STATE Chunk \emph{chunk} $\leftarrow$ getTop(\emph{ST})
   \RETURN \emph{S} $\leftarrow$ getState(\emph{chunk})
\end{algorithmic}
\end{algorithm}
\noindent

Copying can be more memory intensive than trailing, and the intensive memory consumption may introduce non-negligible garbage collection cost, especially for large problems with a substantial number of variables; Schulte has conducted a deep study on this topic in~\cite{comparing}. Nevertheless, it is essential to ensure multiple states are simultaneously available for parallel search, and the concurrent constraint programming system Mozart/Oz~\cite{Oz,ozmodel} adopts copying as one of its restoration techniques.

\subsection{Recomputation} \label{recomp}
 
%In its extreme form, recomputation restores a state by computing from the root of the search tree, using previous generated constraints. Recomputation-based restoration implementsby using the choices stored in chunks. For example, the corresponding pseudocode of  depicted in Algorithm commits the branching constraints in each chunk and then initiates propagation, as described in and implemented in Gecode.
  
Recomputation implements $Record_{recomp}$ to store the constraints that were generated at fix points states and the $Restore_{recomp}$ computes from root state downwards to restore the target state, using the set of the introduced constraints between them. Algorithm~\ref{alg2} depicts the pseudcode for implementing recomputation. In this depiction, the \textbf{for} loop commits constraints in a batch~~\cite{batch} rather than step-wisely as implemented in Gecode.
  
  \begin{algorithm}[h!]
\caption{ $\textit{Restore}_{recomp}$ \label{alg2} }
 \algsetup{linenosize=\tiny}
  \scriptsize
\begin{algorithmic}[1]
   \REQUIRE State \emph{S}, Stack \emph{ST}, Log \emph{log} (ignored)
   \ENSURE State \emph{S}
   \STATE \textbf{delete} \emph{S}
   \STATE \emph{S} $\leftarrow$ getRootState(\emph{ST})
     \STATE \textbf{call} EXPOSE\_OPEN\_CHOICE
      \FOR {\textbf{each} \emph{chunk}  $\in$ \emph{ST} }
         \STATE  \emph{choice} $\leftarrow$ getChoice(\emph{chunk})
          \STATE Commit(\emph{S}, \emph{choice}, \emph{oldAlternative})
       \ENDFOR
   \RETURN \emph{S}  %\emph{\textbf{Propagate}}(\emph{S})
  
\end{algorithmic}
\end{algorithm}
  
Recomputation consumes little memory at the expense of a considerable runtime penalty since computation from scratch is performed whenever a failure occurs. To alleviate, a hybrid scheme can be formed by placing state copies somewhere within the search tree; typical examples are \emph{fixed recomputation} and \emph{adaptive recomputation}. Fixed recomputation places a state copy after every \emph{d} exploration steps, where \emph{d} is a constant called \emph{copying distance}. Fixed recomputation was extended to \emph{adaptive recomputation}: when recomputing from $S_{1}$ to $S_{2}$, it will put an additional state copy in the middle between $S_{1}$ and $S_{2}$ to shorten future recomputation distance. Adaptive recomputation has been demonstrated as a competitive restoration technique~\cite{comparing} and is supported by the Mozart/Oz as well as Gecode systems.

\section{Recollection} \label{recollection}
%\input{recollection}

%As we have reviewed in previous section, recomputation needs to re-execute the propagators' filtering algorithm to compute a visited state. However, this may cause expensive runtimes if the propagators are time-consuming to run and the search process is failure-intensive. Copying-based restoration should be able to alleviate the possible runtime issue, but it would consume more memory than recomputation for the same problem; meanwhile, it can also incur a runtime penalty as a result of copying large states and memory management. Unlike copying, trailing is concerned with data structures within states and records the necessary information to roll back operations when needed; it however does not provide an easy interface to support the attempt of parallel search, and a trail structure that does not use time-stamping mechanism may introduce drastic increase of memory occupation. On the other hand, we can observe that a reason fix point is actually the result of constraint propagation, and the result is reflected as the modification of variable domains. Therefore, it should be feasible to explore state restoration by maintaining the constraint propagation results at the level of variable domains. 

As we have reviewed in the previous section, recomputation needs to re-execute the propagators' filtering algorithm to compute a visited state, while copying stores every reasoned fix point state. Unlike copying, trailing instead is concerned with data structures within states and records the information to roll back performed changes when needed. In this section, we propose an alternative restoration technique that we call \emph{recollection}, which memoizes the propagation updated variable domains to reach a fix point for achieving restoration. By following the presentation paradigm in the previous section, we intend to explain the idea of recollection by describing the implementation of the pair of abstract methods, \emph{Record} and \emph{Restore}.

\subsection{The Record Method.} \label{recordrecollect}

 As illustrated in Algorithm~\ref{alg1}, the \emph{Record} method is invoked after a fix point has been reached. Since the goal of recollection is to memoize propagation affected variable domains for restoration, we should identify the set of variables that were changed during propagation and memorize their domains, as explained in the \textbf{for} loop in Algorithm~\ref{alg5}.

 \begin{center}
 \begin{algorithm}[h!]
 \caption{$\textit{Record}_{\textit{recollect}}$ \label{alg5} }
  \algsetup{linenosize=\tiny}
   \scriptsize
 \begin{algorithmic}[1]
    \REQUIRE State \emph{S}, Choice \emph{choice}, Log \emph{log} (ignored)
    \ENSURE  Chunk \emph{chunk}
     \STATE Domain \emph{doms} $\leftarrow$ $\emptyset$
         \FOR{ \textbf{each}  \emph{var} $\in$ Variables(\emph{S}) }
          \IF  {isChanged(\emph{var})}
              \STATE  \emph{doms} $\leftarrow$ \emph{doms} $\cup$ recordDomain(\emph
              {var})
            \ENDIF
           \ENDFOR
     \RETURN Chunk(\emph{choice}, \emph{doms})
 \end{algorithmic}
 \end{algorithm}
 \end{center}

%Recall that the method  $\textit{Record}$ is invoked in  after a fix point is reached. To support recollection, the method $\textit{Record}_{\textit{recollect}}$ identifies all variables whose domains have changed during the fixpoint computation. To implement domain memoization, it copies all new domains of the changed variables and creates a chunk to store them. 

\subsection{The Restore Method.}  \label{restorerecollection}
 
To restore a state, recollection faces an issue that the required variable domains scatter over the chunks above the target node within the search tree. Therefore, recollection should collect variable domains efficiently for restoration, and we propose two manners: the variable-centered and the chunk-centered.

\paragraph{Variable-centered restoration.} 
 
\begin{algorithm}[h!]
\caption{ $\textit{Restore}_{\textit{variable-centered}}$\label{alg6} }
 \algsetup{linenosize=\tiny} \scriptsize
\begin{algorithmic}[1]
   \REQUIRE State \emph{S}, Stack \emph{ST}, Log \emph{log} (ignored)
   \ENSURE   State \emph{S}'
   \STATE delete \emph{S}
    \STATE \emph{S} $\leftarrow$ getRootState(\emph{ST})   
   \STATE \textbf{call} EXPOSE\_OPEN\_CHOICE
   \FOR {\textbf{each} \emph{var}  $\in$ Variables(\emph{S}) }
\STATE Integer \emph{index} $\leftarrow$ Size(\emph{ST}) - 1
    \STATE  chunk $\leftarrow$ getChunk(\emph{ST}, \emph{index}) \hspace*{1.5mm}/*scan from stack top*/
    \WHILE {Domain(\emph{var}) $\notin$ \emph{chunk}}
        \STATE index $\leftarrow $ index - 1 \hspace*{1.5mm} /*move to next chunk location*/
        \STATE chunk $\leftarrow$ getChunk(\emph{ST}, \emph{index})
     \ENDWHILE
         \STATE  Reconstruct(\emph{var}, \emph{chunk})
    \ENDFOR
    \RETURN \emph{S}
\end{algorithmic}
\end{algorithm}

The variable-centered approach, showing in Algorithm \ref{alg6}, picks one variable \emph{var} at a time and searches the stack \emph{ST} in a top-down direction (moving in the search tree in a bottom up direction!) for the first chunk that contains its domain (Line 7 to 10) and then reconstructs it (Line 11). Suppose the constraint problem imposes  $M$ variables and the number of chunks within the stack is $N$. In the worst case, this approach would conduct $N \times M$ chunk access operations to achieve a restoration; a rather weak propagation problem may approach such a worst-case scenario. On the other hand, most variables can retrieve their domains at the top chunk in the presence of a strong propagation problem and thus the restoration would conduct slightly more than $M$ times chunk access operations.

\begin{algorithm}[h!]
\caption{ $Restore_{chunk-centered}$\label{alg7} }
 \algsetup{linenosize=\tiny}
  \scriptsize
\begin{algorithmic}[1]
   \REQUIRE State \emph{S}, Stack \emph{ST}, Log \emph{log} (ignored)
   \ENSURE   State \emph{S}'
   \STATE delete \emph{S}
   \STATE \emph{S} $\leftarrow$ getRootState(\emph{ST})  
     \STATE \textbf{call} EXPOSE\_OPEN\_CHOICE
\STATE Integer \emph{index} $\leftarrow$ Size(\emph{ST}) - 1
      \WHILE {\emph{index} $\ge$ 0}
\STATE Chunk \emph{chunk} $\leftarrow$ getChunk(\emph{ST}, \emph{index})
       \FOR {\textbf{each} \emph{var} $\in$ Variables(\emph{chunk}) }
       \IF { \emph{var} has not been reconstructed}
              \STATE   Reconstruct(\emph{var}, \emph{chunk})
              \ENDIF
              \ENDFOR
        \STATE \emph{index} $\leftarrow$ \emph{index} - 1
    \ENDWHILE
    \RETURN \emph{S}
\end{algorithmic}
\end{algorithm}

\paragraph{Chunk-centered restoration.}

By contrast, the chunk-centered approach, showing in Algorithm \ref{alg7}, scans \emph{ST} in a top-down manner (moving bottom-up in the search tree) and keeps track of reconstructed domains. Within each chunk, all memoized variables are scanned and a variable domain will be reconstructed if it has not been reconstructed yet (Lines 6--14). This query scheme requires to access the stack once during a restoration, regardless of whether the problem exhibits weak or strong propagation, and it has a slight runtime advantage over variable-centered restoration according to our experimental results. The experiments of the next section have been conducted using the chunk-centered restoration; we introduce an index to facilitate the recognition of reconstructed variables.

\subsection{Variants} \label{variant}

Our discussion on recollection so far assumes that a single state is maintained at the search tree root and thus restoration begins from scratch. We observer that recollection alone usually incurs a significant runtime penalty, especially for the problems that create a deep search tree. This issue attributes to the process of searching chunks when recollection intends to restore a state. To alleviate the penalty, we extend recollection to the variants of \emph{fixed recollection} and \emph{adaptive recollection}, analogous to the recomputation variants; they place state copies somewhere in the search in the way that have been explained in Section~\ref{recomp}.

A recent studied restoration technique in~\cite{states} intends to combine trailing, copying and recomputation to construct a hybrid architecture. In addition to variable domains, this hybrid scheme is also concerned about propagators and propagator states as well as the dependencies between variables and propagators. On the other hand, recollection concentrates on variable domains and is transparent with other detail. Meanwhile, recollection also can be orthogonal with the implementation of search.

\subsection{Implementation Issues}
 
 Gecode is a system that develops computation space (space for short) as its central concept. A computation space encapsulates computations and is home to variables, propagators and branchers. Computation spaces are first-class structures that can be copied and programmed to construct search engines using its provided operations, a computation space corresponds a node in the search tree. In this sub-section, we discuss three key issues to implement recollection within Gecode system.

\subsubsection{Memory Management}

The implementation of method $\textit{Record}_{\textit{recollect}}$ consumes memory, but Gecode's \emph{memory manager} is centered on spaces, while chunks live most naturally outside spaces. Hence, we should provide a memory management policy to support the memorized variable domains. In our prototype, we investigated two options to address the memory management issue: (1). allocate memory dynamically for recorded each variable; (2). calculate the expected total memory size first and then allocate once. The allocated memory is freed when its chunk is popped from the stack. Our experiments reveal that approach (1) is marginally more runtime efficient for problem with weak propagation, while approach (2) is more suitable for problems of many variables with strong propagation. In our experimental prototype~\cite{prototype}, we can switch between the two alternatives by setting a compile-time flag. Note that we use approach (1) in all experiments reported in the next section.

\subsubsection{Domain Change Detection}

A key implementation challenge of the method $\textit{Record}_{\textit{recollect}}$ in Section~\ref{recordrecollect} is identifying the set of variables that were updated to reason each fix point. Fortunately, our chosen recollection implementation platform, Gecode, developed an abstraction called \emph{advisor}~\cite{advisor} to optimize constraint propagation. We make use of the instantiated advisor structures to recognize the changed variables. For detail of the technique to achieve this goal, please refer to the source code of our published recollection implementation~\cite{prototype}.

\subsubsection{Variable Reconstruction}

For a variable domain, the removal of values by constraint propagation may break it into multiple intervals. In Gecode, each interval is implemented as a \emph{range} data structure and multiple ranges within a variable implementation will be chained as a list. In addition to updating the values represented on each range, recollection may be required to adjust the current chain length as well. We developed two methods to achieve this goal. 

The first option simply destroys the current chain and then rebuilds another one. This method is straightforward to implement, but frequent chain destroying may introduce intensive garbage collection for the system, especially when dealing with long chains. Therefore, this alternative should perform better to deal with short and medium length chains. By contrast, the second approach either trims or extends the current chain to the new length. In our prototype, we have implemented both alternatives and they are configurable through a compile-time flag. For the experiments in evaluation section, we employ the second approach.

%
%\section{Implementation} \label{implementation}
%\%input{implementation}
%
\section{Evaluation} \label{evaluation}

This section empirically evaluates recollection over a set of benchmark problems: Section~\ref{sub1} describes the overall evaluation setting; Section~\ref{sub2} compares recomputation with recollection and Section~\ref{sub3} intends to extend comparison to other restoration techniques.

\subsection{Configurations}\label{sub1}

We used a PC system that is equipped with Intel Core 2 Quad processor Q9550, running the Ubuntu operating system 12.04 in a 32-bit mode with four Gigabyte main memory. We built our prototype~\cite{prototype} on top of Gecode version 3.7.3~\cite{Gecode}, which also served as the reference instance for comparison; the source code was compiled by G++ version 4.6.3. Each collected runtime\footnote{We take wall clock time in this work.} value is an arithmetic mean of 20 runs with a variation coefficient less than 2\%; memory measurements is the \emph{peak} amount of the memory occupation. 

As benchmarks, we used finite domain integer and Boolean problems. They were selected to cover multiple constraints, spawn a varying number of propagators and impose different propagation intensity. Meanwhile, they cover first, all and best (branch-and-bound) solution search. We limit ourselves to the problems included in the Gecode repository and stick to the configuration of propagation consistency level and branching strategy of the original scripts. 

\begin{table}[h!]
 \begin{center}
 \begin{tabular}{l c r r r r r}
 \toprule
 Problem & Sols & Propagators & Propagations & Nodes     & Failures  & Depth   \\
 \midrule
 Queens(100) & one&14,850 & 16,821 & 138 & 22 & 96 \\
 Queens-S(100) & one&3 & 428 & 138 & 22 & 96 \\
 Magic-Square(5) & one& 15 & 2,292,251 & 144,471 & 72,227 & 33 \\
 Sport-League(22) & one & 1,199 & 207,066 & 2,273 & 1,035 & 249 \\
 Black-Hole & one&742 & 986,542 & 5,284 & 2,631 & 47 \\
 BIBD & one &9,693 & 912,464 & 2,625 & 1,306 & 968 \\
 %Ortho-Latin & one&133 & 356,768 & 4,405 & 2,188 & 46 \\
 Knight(22) & one&1 & 74,610 & 40,184 & 19,877 & 451 \\
 Pentominoes & one &81 & 6998 & 143 & 64 & 27 \\
 Alpha & all & 21 & 136,179 & 14,871 & 7,435 & 49\\
 Langford-Num & all & 37 & 22243 & 303 & 149 & 17 \\
 Golomb-Ruler(10) &optimal& 39 & 2,760,799 & 39,875 & 19,928 & 33 \\
 Ind-Set & optimal &21 & 101,317 & 29,849 & 14,895 & 40 \\
 \bottomrule
\end{tabular}
\end{center}
 \caption{Characteristic of Benchmark Problem Search Trees}
 \label{treeinfo}
\end{table}

The set of selected benchmark problems are: the Queens problem modelled by either a quadratic number of disequality constraints or three global constraints that generalize \emph{all-different}; the magic-square puzzle of size 5; a sport league problem with 22 teams; the black hole patience game; Balanced Incomplete Block Design (BIBD), the knights tour problem of size 22; the Pentominoes problem; the Alpha crypto-arithmetic puzzle; the Langford's number problem with 3 by 9 values and; Golomb-Ruler problem of size 10 and the problem of independent sets in graph (Ind-Set). Table~\ref{treeinfo} lists the characteristics of these problems, where the \emph{propagations} are the numbers collected when using adaptive recomputation for restoration with default argument settings. For the original scripts, refer to the Gecode distribution in~\cite{Gecode}.

\subsection{Recomputation and Recollection} \label{sub2}

The proposal of recollection was motivated by recomputation in a sense of skipping the re-execution of constraint propagation, hence the foremost performance comparison should be with recomputation. As explained in Section~\ref{restoration}, recomputation can derive variants by combining with copying, and adaptive recomputation generally exhibits superior runtime performance compared with other recomputation schemes~\cite{comparing}. Similarly, adaptive recollection exposes the most runtime competitive recollection scheme. We therefore first focus on a direct comparison between adaptive recomputation and adaptive recollection, fixing the copying distance to \emph{eight} in both cases.

\begin{table}[h!]
\begin{center}
\begin{tabular}{l r r r r r r }
%\hline
\toprule
 & \multicolumn{2}{c}{Recomputation} & & & \multicolumn{2}{c}{Recollection} \\
\cline{2-3} \cline{6-7}
   Problems & Time(ms) & Mem(KB) & & & Time (ms) & Mem(KB)  \\
\midrule

 Queens(100) & 16 & 4,301 & & & 15 & 4,663 \\
 Queens-S(100) & 1  & 240 & & & 2 & 602 \\
 Magic-Square(5) & 579 & 63 & & & 653 & 73\\
 Sport-League(22) & 352 & 7,710 & & & 331 & 7,937\\
 Black-Hole & 535 & 1,927 & & & 508 &  1998\\
 BIBD & 573 & 4,678 & & & 575 & 4784\\
 %Ortho-Latin & 211 & 1319 & & & 191 & 1377\\
 Knights(22) &1,858 & 4,460 & & & 1,704 & 4,592\\
 Pentominoes & 20 & 1,158 & & & 19 & 1,173 \\
 Alpha & 55 & 45 &&& 66 & 50 \\
 Langford-Number & 13 & 132 & & & 13 & 135 \\
 Golomb-Ruler(10)& 556 & 69 & & & 547 & 70 \\
 Ind-Set & 58 & 41 & & & 68 & 43 \\

\bottomrule
\end{tabular}
\end{center}
 \caption{Comparison  Adaptive Recomputation and Adaptive Recollection}
 \label{firstComparison}
\end{table}

Table~\ref{firstComparison} depicts the experimental results, which demonstrates that neither recomputation nor recollection can impose a consistent performance advantage over the other for solving all problems. Recollection hardly improves the runtime of the problems with shallow search trees and limited number of failures such as Pentomonies and Langford-Number; or even leads to an inferior runtime, as in Alpha and Magic Squares. Nevertheless, recollection can be runtime competitive for finite domain integer problems with deep search trees and intensive search failures such as Sport-League, Golomb-Ruler and Knights. On these problems, recollection is able to make an runtime improvement by investing a small amount of additional memory than adaptive recomputation. 

Boolean problems can hardly benefit from recollection, even though a Boolean problem would explore a rather deep search tree (BIBD) and incur intensive failures (Ind-Set). This is mainly because that a Boolean variable contains at most two values; recollection memoized Boolean domains are not dense enough to compete with the recomputation via re-runing propagation algorithms. 

The measurements in Table~\ref{firstComparison} was conducted with a specific copying distance (\emph{eight}), and this may impose a concern that the fixed copying distance may conclude skewed results. To dispel this concern, we ran Sport-League and Knights problem in adaptive recomputation and adaptive recollection respectively over a range of copying distance values, and Table~\ref{adaptivity} illustrates the collected runtime measurements.

\begin{table}[h!]
\begin{center}
\begin{tabular}{l c c c c c c c c c c c }
%\hline
\toprule
 & \multicolumn{11}{c}{Copying Distance $(d)$ } \\
\cline{2-12}
    & $d=1$  & $d=3$ & $d=5$ & $d=10$ & $d=20$ & $d=40$ & $d=80$ & $d=160$ & $d=320$  \\
    \midrule
 SportLeague(\emph{recomp})  & 337& 330  &341 & 350&351 & 355& 359& 360 & 359 \\
 SportLeague(\emph{recoll})  & 336& 325  &326 & 330&329 & 333& 334& 336 & 335  \\
   \textbf{Time}$\Delta$(ms) & 1  & 5 & 15 & 20 & 22 & 22 & 25 & 24 & 24\\
 \hdashline
 Knights(\emph{recomp})  &1598 &1737 &1830&1856&1855&1868&1872&1855&1864\\
 Knights(\emph{recoll}) &1589 & 1646  & 1695 & 1712 & 1711 & 1703 & 1697&1700 &1714\\
 \textbf{Time} $\Delta$(ms) & 9 & 91 & 135 & 144  & 144 & 165 & 175 & 155 & 150 \\
\bottomrule
\end{tabular}
\end{center}
 \caption{Sport-League and Knight run over a range of copying distances}
 \label{adaptivity}
\end{table}

Table~\ref{adaptivity} shows that adaptive recollection is able to adjust quickly to converge to a small runtime interval, even though the copying distance is set to a large value. This observation indicates that the setting of copying distance is not as significant as one may have imagined, which confirms and generalizes the corresponding original observation reported on adaptive recomputation in~\cite{comparing}. As observed from the table, the runtime difference between recomputation and recollection initially grows as copying distance increases and then shrinks somewhat after reaching a peak performance gap (at \emph{d}=80 in both cases); afterwards, it stays almost stable with the further increase of the copying distance.

\subsection{Extended Comparison} \label{sub3}

In addition to recomputation, the evaluation would be more thorough if it includes copying and trailing. Fortunately, copying-based restoration can be easily obtained by setting the copying distance to \emph{one} in Gecode. As for trailing, it is absent in current Gecode system. To implement a trailing-based restoration, we are expected to clear many techniques choices such as trail structure, and these factors are key to the performance of a trailing-based system. Meanwhile, Gecode centers on copying with recomputation and it has developed intensive techniques to optimize its underlying restoration. In such a circumstance, a concern about fair comparison would come out. Alternatively, another way is to employ a set of trailing-based systems to carry out platform-crossing empirical evaluation. Such a work would indeed be a significant contribution, and Schulte has conducted a similar investigation in~\cite{comparing}. However, this exceeds our resource in this paper, the goal of which is to propose an alternative restoration technique. Instead, we take the systematic platform-crossing study as one of our main future research works. Nevertheless, we intend to illuminate the comparison with other restoration techniques as illustrated in Table~\ref{connect}.  

\begin{table}[h!]
\begin{center}
\begin{tabular}{l c c c c c c c c }
%\hline
\toprule
 & \multicolumn{2}{c}{Copying} & & \multicolumn{2}{c}{Recomputation} & & \multicolumn{2}{c}{Recollection} \\
\cline{2-3} \cline{5-6 }\cline{8-9}
   Problems & Time(ms) & Mem(KB) & & Time(ms) & Mem(KB) & & Time(ms) & Mem(KB)  \\
\midrule

 Queens(100) & 39 & 26076 &  & 16 & 4301  &  & 15 & 4663\\
 Queens-S(100)& 2 &1662  &  & 1 & 240  &  & 2 & 602\\
 Queens(200) & 3837 & 170669 &  & 4298 & 6224  &  & 4626 & 9027\\
 Queens-S(200)& 1996 &8560  &  & 2171 & 1066  &  & 2479 & 3866\\
 Alpha & 51 & 54 &  & 55 & 45  &  & 66 &50 \\
 Magic-Square & 487 & 100 &  & 579 & 58 &   & 652   & 67\\
  Knights(18) & 31 & 11271 &  & 20 & 1596  &  & 19 & 1681\\
 Knights(22) & 1598 & 30159 &  & 1858 & 4460  &  & 1704 & 4592\\
 Golomber-Ruler & 469 & 77 &  & 550 & 61 &   & 550   & 63\\

\bottomrule
\end{tabular}
\end{center}
 \caption{Comparison with other restoration techniques}
 \label{connect}
\end{table}

The table reveals that recollection can consume much less memory than copying as one can expect for most problems. For runtime, recollection generally cannot compete with copying except on few cases. In~\cite{states}, it implemented a simplified trailing in Gecode to compare with other restoration strategies. Its conducted evaluation demonstrates that copying is slightly preferable if problems impose strong propagation, while trailing benefits a lot in weak propagation problems. This finding may entail a promising comparison between recollection and trailing since recollection has developed methods (variable-center restoration in Section~\ref{restorerecollection}) to aim for efficient restoration in strong propagation problems.

%Schulte~\cite{comparing} observes that copying together with adaptive recomputation can outperform trailing-based system for large problems with deep search trees (Queens-100 and Knights-18). Queens-100 and Knights-18. For runtime, recollection does not outperformcopying on small or medium size problems; however, it cuts the runtime almost in half on large problems (Queens-100 and Knights-18).Recollection futher  improves the runtime on the two benchmark problems Queens-100 and Knights-18 problems.

%\begin{table}[h!]
%\begin{center}
%\begin{tabular}{l c c c c c c c c }
%%\hline
%\toprule
% & \multicolumn{2}{c}{Copying} & & \multicolumn{2}{c}{Recomputation} & & \multicolumn{2}{c}{Recollection} \\
%\cline{2-3} \cline{5-6 }\cline{8-9}
%   Problems & Time(ms) & Mem(KB) & & Time(ms) & Mem(KB) & & Time(ms) & Mem(KB)  \\
%\midrule
%
% Alpha & 51 & 54 &  & 55 & 45  &  & 66 &50 \\
% Queens(10) & 26 & 77 &  & 34 & 53  &  & 37 & 55\\
% Queens-S(10)& 17 &41  &  & 21 & 29  &  & 25 & 31\\
% Queens(100)& \textbf{39} & \textbf{26076} &  & \textbf{16} & \textbf{4301}  &  & \textbf{15} &\textbf{4663} \\
% Queens-S(100)& 3 & 1662 &  & 1 & 240 &   & 2 & 602 \\
% Magic-Seq & 51 & 4358 &  & 51 & 4358 &   & 51   & 4361\\
%
%
%\bottomrule
%\end{tabular}
%\end{center}
% \caption{Comparison with other restoration techniques}
% \label{connect}
%\end{table}

\section{Conclusion} \label{conclusion}
%\input{conclusion}
%We felt that recollection is a simple state restoration technique that needed to be tried, to complete the range of possibilities. As opposed to trailing, it proceeds top-down and only needs to recollect a domain once per fix point. As opposed to space copying, it only memorizes the domains of changed variables. As opposed to space recomputation, it does not need to re-execute propagator implemented filtering algorithms.
%
%
%The Gecode system was especially suitable for implementing recollection because of its advisor facility~\cite{advisor}. The evaluation of our prototypicial implementation demonstrates that pure recollection is able to offer a prominent memory improvement comparing with copying (at the cost of runtime) and a clear runtime improvement compared to pure recomputation (at the cost of memory). The comparison of adaptive recollection with adaptive recomputation shows that that marginal runtime improvements are possible at the cost of a very moderate memory overhead. The current implementation is done on top of a system designed around a different restoration philosophy, and therefore we conjecture that more significant runtime improvements can be expected if one would develop a full-fledged constraint programming system from scratch, with recollection as its central restoration philosophy.

We proposed a new top-down state restoration technique that we call \emph{recollection} and present it in an exposition using the abstract methods $\textit{Record}$ and $\textit{Restore}$. Unlike copying, recollection exhibits a finer granularity and only memoizes the variable domains that were changed during each fix point computation; compared with recomputation, it would not re-run the propagators' built-in filtering algorithms for restoration; as opposed to trailing, recollection approaches in a top-down direction within the search tree, and it is orthogonal with search.

%  Compared to the
%existing top-down technique of copying, recollection exhibits a finer
%granularity and only copies the domains of variables that have changed
%during each fixpoint computation. Compared to the top-down restoration
%technique of recomputation, it avoids re-running of the propagator
%filtering algorithms.  As opposed to the bottom-up restoration
%technique of trailing, recollection avoids repetitive variable
%restorations within the same fixpoint and is not limited to
%chronological backtracking.

Our experimental evaluation reveals that recollection has the opportunity to improve runtime against adaptive recomputation on integer problems with deep search trees and intensive search failures, at the expense of moderate memory investment. Our current prototype was built on Gecode system and has investigated several implementation alternatives to develop this idea, and we cautiously conjecture more significant performance improvements if one takes further efforts to explore more aggressive techniques to optimize recollection implementation. We hope that the presented recollection technique is significant enough to be added to the set of tools available to designers and implementers of constraint programming systems. Exciting future research topics can be exploring parallel and non-chronological backtracking search by utilizing recollection as the underlying restoration strategy. Additionally, an in-depth comprehensive system-crossing performance study on various restoration techniques would be another significant contribution.

\subsection*{Acknowledgements}

We are very grateful to Christian Schulte and Guido Tack for unfailingly resourceful answers to our pestering queries and their (cautious) words of encouragement.

 \bibliographystyle{plain}

 \bibliography{references}

\end{document}